\documentclass[a4paper,8pt]{article}

\begin{document}

\newcommand{\munu}{^\mu_{\phantom{\mu}\nu}}
\makeatletter
\@addtoreset{equation}{section}
\def\theequation{\thesection.\arabic{equation}}
\makeatother

\title{Violation of time symmetry}
\author{Koji Nagata\\
{\it Department of Physics,}\\
{\it Korea Advanced Institute of Science and 
Technology,}\\
{\it Taejon 305-701, Korea}\\
{E-mail:nagata@kaist.ac.kr}\\
{TEL:+82-42-869-2596}\\
{FAX:+82-42-869-2510}}

\date{}
\maketitle

\begin{abstract}
Recently, [{arXiv:0810.3134}] is accepted and published.
We would like to study
the relation between a local realistic theory and 
commutativity of observables in 
a finite-dimensional space.
We would like to conjecture that a realistic theory of the Bell type 
(a local realistic theory)
for events would not imply 
the commutative algebraic structure into 
the set of all observables
if all experimental events
would be reproduced by a local realistic theory.
We would like to suggest that a violation of Bell locality would be derived 
within a realistic theory of the Kochen-Specker type 
within quantum events.
\end{abstract}



\newpage

\section{Introduction}
Recently, \cite{NagataNakamura} is accepted and published.
As is well known, quantum observables do not commute generally 
in the Hilbert space formalism of the 
quantum theory \cite{bib:Redhead,bib:Peres}.
It is one of nonclassical features of the quantum theory.
It may be said as noncommutativity.
The equivalence between noncommutativity 
and the Kochen-Specker (KS) theorem \cite{bib:KS} is 
shown in Refs.~\cite{bib:Malley1,bib:Malley2}.

Other nonclassical feature of the quantum theory is 
a violation of Bell locality \cite{bib:Bell}.
It may be said as Bell nonlocality.
This feature is derived by the inner product machinery of 
the Hilbert space \cite{bib:Zukowski0}.
The norm of Hermitian operators is generated 
by the inner product.
A set of inner products violates the inequalities imposed 
by Bell locality.
Similar situation occurs 
when entanglement witness inequalities are violated by 
a set of inner products.
In this case, the inequalities are derived by the assumption that
the system is described by separable states.
We see that both (entanglement and Bell nonlocality) of 
the mathematical derivation 
have similar machinery 
in the Hilbert space formalism of the quantum theory.

It is shown 
that several pure entangled states must violate some Bell 
inequality \cite{bib:Gisin,bib:Zukowski1,bib:Wieslaw}.
There is an entangled mixed state which satisfies 
all Bell inequalities (cf. Ref.~\cite{bib:Werner1}).
Such a state should violate some entanglement witness inequalities.
There is a pure entangled state in 
the mixture of pure states constituting such a mixed state
if the mixed state in question is an entangled state. 
Otherwise, the mixed state is written as 
a convex sum over separable states.
We can see that the notion of entanglement 
of a mixed state is closely 
related with the notion of Bell nonlocality and 
directly with entanglement witnesses.

It is discussed at length that 
entanglement is one of reasons why 
various quantum information 
processes are possible \cite{bib:Nielsen,bib:Galindo}.

It is suggested \cite{bib:Fine2} that a violation of 
a Bell inequality implies noncommutativity.
It is conjectured that the converse 
proposition is also true via 
the existence of joint distributions.
`Joint distributions are well defined 
only for commuting observables' \cite{bib:Fine}.
This seems reasonable. 

It is also
suggested \cite{bib:Malley3,Mall} that all 
quantum observables would 
commute simultaneously if 
we would accept a realistic theory of the Bell type for quantum events,
provided that all quantum events, 
including every quantum state and every observable 
(including every projector),
would be reproduced by a realistic theory of the Bell type 
for quantum events.
It seems that this indeed provides very important location that 
we should discuss about this open problem.

In this paper, we would like to study
the relation between a local realistic theory and 
commutativity of observables in 
a finite-dimensional space.
We would like to conjecture that a realistic theory of the Bell type 
(a local realistic theory)
for events would not imply 
the commutative algebraic structure into 
the set of all observables
if all experimental events
would be reproduced by a local realistic theory.
And we would like to suggest 
that a violation of Bell locality would be derived 
within a realistic theory of the KS type 
within quantum events.

We would like to hope that 
our discussion might provide further information for 
people who further consider the open problem.

Our thesis is organized as follows.
In Sec.~\ref{Notation},
we would like to provide notation and preparations for this paper.
In Sec.~\ref{Bellthe},
we would like to review the Bell theorem.
In Sec.~\ref{axiom},
we would like to study a theorem concerning 
the relation between a local realistic theory and commutativity of observables.
In Sec.~\ref{MFinee},
we would like to study
the relation between the Bell theorem and commutativity of observables.
In Sec.~\ref{exam},
we would like to suggest that a violation of Bell locality would be derived 
within a realistic theory of the KS type 
within quantum events.
Section \ref{conclusion} concludes this paper.

\section{Notation and preparations}\label{Notation}

We consider a finite-dimensional space $H$.
Let ${\bf R}$ denote the reals where $\pm\infty\not\in {\bf R}$.
We assume that every result of measurements lies in ${\bf R}$.
We assume that every time $t$ lies in ${\bf R}$.
Let ${O}$ be the set of all observables in a space $H$.
Let ${T}$ be the set of all states in the space $H$.
We define a notation $\theta(t)$ which represents 
one result of measurements in a time $t$.
We assume that 
measurement of 
an observable $A$ in a time $t$ 
for a physical system in a state $\psi(t)$ yields a 
value $\theta(A,t)\in {\bf R}$.
We define $\Delta$ as any subset of the reals ${\bf R}$.
We define $\chi_{\Delta}(x), (x\in{\bf R})$ as
the characteristic function.
We assume that there is a classical probability space.
It is written as $(\Omega,\Sigma,\mu_{\psi(t)})$.
$\Omega$ is a nonempty space. 
$\Sigma$ is a $\sigma$-algebra of subsets of $\Omega$. 
$\mu_{\psi(t)}$ is a $\sigma$-additive normalized measure on $\Sigma$ 
such that $\mu_{\psi(t)}(\Omega)=1$.
The letter written as subscript $\psi(t)$ means that
the probability measure is determined uniquely
when a state $\psi(t)$ is specified.

We introduce measurable functions 
(classical random variables) onto $\Omega$
($f: \Omega \mapsto {\bf R}$).
The measurable function is written as $f_A(\omega_t)$
for an observable $A\in {O}$.
Here $\omega_t\in\Omega$ is 
an element with respect to a time $t$.
Let $S$ be $\{\pm 1\}$.

We would like to consider the following propositions:

{\bf Proposition:} R ({\it a 
realistic interpretation of a physical theory}),

A measurable function $f_A(\omega_t)$ exists for 
every observable $A$ in ${O}$ and for every time $t$.

{\bf Proposition:} D ({\it the probability distribution rule}),
\begin{eqnarray}
\mu_{\psi(t)}(f^{-1}_{A}(\Delta))
={\rm Prob}(\Delta)_{\theta(A,t)}^{\psi(t)}.
\end{eqnarray}
The symbol $(\Delta)_{\theta(A,t)}^{\psi(t)}$ 
denotes the following proposition:
$\theta(A,t)$ lies in $\Delta$ if the system is 
described by a state $\psi(t)$.
The symbol ``${\rm Prob}$'' denotes the probability that
the proposition $(\Delta)_{\theta(A,t)}^{\psi(t)}$ holds.

{\bf Proposition:} L ({\it the Bell locality}),
\begin{eqnarray}
\omega_{t_1}=\omega_{t_2}=\omega_{t}
\end{eqnarray}
for $t_1\neq t_2$.

{\bf Proposition:} M ({\it measurement outcome}),
\begin{eqnarray}
f_A(\omega_t)\in S.\label{MO}
\end{eqnarray}

We would like to review the following:

{\bf Lemma:}\cite{bib:Nagata2}
If
\begin{eqnarray}
\Vert\{\psi(t), A\}\Vert&:=&\sum_{y\in S} 
{\rm Prob}(\{y\})_{\theta(A,t)}^{\psi(t)}y,\nonumber\\
E_{\psi(t)}(A)&:=
&\int_{\omega_t\in \Omega}\mu_{\psi(t)}({\rm d}\omega_t)
f_{A}(\omega_t),\nonumber
\end{eqnarray}
then 
\begin{eqnarray}
{\rm R}\wedge{\rm D}\wedge{\rm M}
\Rightarrow
\Vert\{\psi(t), A\}\Vert=E_{\psi(t)}(A).\label{QMHV}
\end{eqnarray}

{\it Proof:}
Note
\begin{eqnarray}
&&\omega_t\in f^{-1}_{A}(\{y\})\Leftrightarrow
f_{A}(\omega_t)\in \{y\}\Leftrightarrow
y=f_{A}(\omega_t),\nonumber\\
&&\int_{\omega_t\in f^{-1}_{A}(\{y\})}
\frac{\mu_{\psi(t)}({\rm d}\omega_t)}{\mu_{\psi(t)}
(f^{-1}_{A}(\{y\}))}=1,\nonumber\\
&&y\neq y'\Rightarrow 
f^{-1}_{A}(\{y\})\cap f^{-1}_{A}(\{y'\})=\phi.
\end{eqnarray}
Hence we have
\begin{eqnarray}
&&\Vert\{\psi(t), A\}\Vert=
\sum_{y\in S}{\rm Prob}(\{y\})_{\theta(A,t)}^{\psi(t)}y
=\sum_{y\in{\bf R}} 
{\rm Prob}(\{y\})_{\theta(A,t)}^{\psi(t)}y\nonumber\\
&&=\sum_{y\in{\bf R}}\mu_{\psi(t)}(f^{-1}_{A}(\{y\}))y\nonumber\\
&&=\sum_{y\in{\bf R}}
\mu_{\psi(t)}(f^{-1}_{A}(\{y\}))y
\times \int_{\omega_t\in f^{-1}_{A}(\{y\})}
\frac{\mu_{\psi(t)}({\rm d}\omega_t)}{\mu_{\psi(t)}
(f^{-1}_{A}(\{y\}))}\nonumber\\
&&=\sum_{y\in{\bf R}}\int_{\omega_t\in f^{-1}_{A}(\{y\})}
\mu_{\psi(t)}(f^{-1}_{A}(\{y\}))
\times 
\frac{\mu_{\psi(t)}({\rm d}\omega_t)}{\mu_{\psi(t)}
(f^{-1}_{A}(\{y\}))}f_{A}(\omega_t)\nonumber\\
&&=\int_{\omega_t\in \Omega}\mu_{\psi(t)}({\rm d}\omega_t)
f_{A}(\omega_t)=E_{\psi(t)}(A).
\end{eqnarray}
QED.

The probability measure $\mu_{\psi(t)}$ may be
chosen such that the following equation
would be valid if we assign the truth value ``1'' for 
Proposition R, Proposition D, and Proposition M, simultaneously:
\begin{eqnarray}
\Vert\{\psi(t), A\}\Vert=
\int_{\omega_t\in\Omega}\mu_{\psi(t)}({\rm d}\omega_t)
f_A(\omega_t)
\end{eqnarray}
for every observable $A$ in ${O}$.

{\bf Definition:} ({\it an observable with respect to commutator}),
\begin{eqnarray}
F(A, B):=|[A, B]|^2
\end{eqnarray}
for every pair of observables $A$ and $B$.

\section{The Bell theorem}\label{Bellthe}

In this section, we would like to review the Bell theorem:

{\bf Theorem:} ({\it the Bell inequality}),
\begin{eqnarray}
&&{\rm R}\wedge{\rm D}\wedge{\rm L}\wedge{\rm M}\nonumber\\
&&\Rightarrow
\Vert\{\psi(t), \sigma^1_x\sigma^2_x\}\Vert
-\Vert\{\psi(t), \sigma^1_x\sigma^2_y\}\Vert
+\Vert\{\psi(t), \sigma^1_y\sigma^2_x\}\Vert
+\Vert\{\psi(t), \sigma^1_y\sigma^2_y\}\Vert\leq 2.
\nonumber\\
\end{eqnarray}

{\it Proof:}
Let $x, y$ be real numbers with
$x, y\in S$.
Then we have
\begin{eqnarray}
(xx-xy+yx+yy)=\pm 2.\label{CHSH}
\end{eqnarray}
Proposition R and Proposition M imply 
\begin{eqnarray}
f_{\sigma^1_x}(\omega_{t_1})= \pm1,~
f_{\sigma^1_y}(\omega_{t_2})= \pm1,~
f_{\sigma^2_x}(\omega_{t_1})= \pm1,~
f_{\sigma^2_y}(\omega_{t_2})= \pm1.
\end{eqnarray}
Hence the condition (\ref{CHSH}) implies
\begin{eqnarray}
&&U(\omega_{t_1}, \omega_{t_2})
:=
f_{\sigma^1_x}(\omega_{t_1})f_{\sigma^2_x}(\omega_{t_1})
-f_{\sigma^1_x}(\omega_{t_1})f_{\sigma^2_y}(\omega_{t_2})
+f_{\sigma^1_y}(\omega_{t_2})f_{\sigma^2_x}(\omega_{t_1})
+f_{\sigma^1_y}(\omega_{t_2})f_{\sigma^2_y}(\omega_{t_2})\nonumber\\
&&\Rightarrow U(\omega_{t_1}, \omega_{t_2})= \pm2.
\end{eqnarray}
Proposition L implies $U(\omega_{t_1}, \omega_{t_2})
=U(\omega_{t}, \omega_{t})$.
Thus,
\begin{eqnarray}
&&U(\omega_{t}, \omega_{t})
=f_{\sigma^1_x}(\omega_{t})f_{\sigma^2_x}(\omega_{t})
-f_{\sigma^1_x}(\omega_{t})f_{\sigma^2_y}(\omega_{t})
+f_{\sigma^1_y}(\omega_{t})f_{\sigma^2_x}(\omega_{t})
+f_{\sigma^1_y}(\omega_{t})f_{\sigma^2_y}(\omega_{t})=\pm 2.\nonumber\\
\end{eqnarray}
We define $V(\omega_t)$ as $V(\omega_t):=U(\omega_{t}, \omega_{t})$.
We see the following implication:
\begin{eqnarray}
&&V(\omega_t)
=f_{\sigma^1_x}(\omega_t)f_{\sigma^2_x}(\omega_t)
-f_{\sigma^1_x}(\omega_t)f_{\sigma^2_y}(\omega_t)
+f_{\sigma^1_y}(\omega_t)f_{\sigma^2_x}(\omega_t)
+f_{\sigma^1_y}(\omega_t)f_{\sigma^2_y}(\omega_t)\nonumber\\
&&\Rightarrow V(\omega_t)= \pm2.
\end{eqnarray}
Hence we have
\begin{eqnarray}
\int_{\omega_t\in\Omega}\mu_{\psi(t)}({\rm d}\omega_t)
V(\omega_t)\leq 2.
\end{eqnarray}
Proposition R, Proposition D, and Proposition M  imply (cf. (\ref{QMHV}).)
\begin{eqnarray}
\Vert\{\psi(t), \sigma^1_x\sigma^2_x\}\Vert
-\Vert\{\psi(t), \sigma^1_x\sigma^2_y\}\Vert
+\Vert\{\psi(t), \sigma^1_y\sigma^2_x\}\Vert
+\Vert\{\psi(t), \sigma^1_y\sigma^2_y\}\Vert
\leq 2.\label{Bell}
\end{eqnarray}
QED.

A violation of the inequality (\ref{Bell}) implies 
that 
we cannot assign the truth value ``1'' for 
Proposition R,
Proposition D, 
Proposition L, and
Proposition M, simultaneously,
in a state $\psi(t)$.

In what follows, we would like to accept the quantum theory.
We may assume that $\Vert\{\psi(t), A\}\Vert
={\rm tr}[\psi(t)A]$ for every $A\in O$.

Let $\psi(t)$ be 
\begin{eqnarray}
\psi(t)=\frac{1}{\sqrt{2}}
(|+_1\rangle|+_2\rangle+e^{i\pi/4}|-_1\rangle|-_2\rangle).
\end{eqnarray}
We may assume that
\begin{eqnarray}
{\rm tr}[\psi(t)\sigma^1_x\sigma^2_x]+{\rm tr}[\psi(t)\sigma^1_y\sigma^2_y]
=\sqrt{2},~
-{\rm tr}[\psi(t)\sigma^1_x\sigma^2_y]
+{\rm tr}[\psi(t)\sigma^1_y\sigma^2_x]
=\sqrt{2}.
\end{eqnarray}
Then we have
\begin{eqnarray}
{\rm tr}[\psi(t)\sigma^1_x\sigma^2_x]
-{\rm tr}[\psi(t)\sigma^1_x\sigma^2_y]
+{\rm tr}[\psi(t)\sigma^1_y\sigma^2_x]
+{\rm tr}[\psi(t)\sigma^1_y\sigma^2_y]=2\sqrt{2}.
\end{eqnarray}
Therefore the Bell theorem would be true if 
we would accept the quantum 
theory.

\section{A local realistic commuting observables theory}\label{axiom}

In this section, we would like to study the following theorem:

{\bf Theorem:} 
\begin{eqnarray}
&&\forall t\in{\bf R}, ~\forall \psi(t)\in
T, ~\forall A, B\in O : \nonumber\\
&&\left[
{\rm R}\wedge{\rm D}\wedge{\rm L}\wedge{\rm M}
\wedge 
\left[
\Vert \{\psi(t), F(A, B)\}\Vert=0
\right]
\right]
\Rightarrow 
\bot.
\end{eqnarray}

{\it Proof:}
Proposition R, Proposition D, and Proposition M imply
\begin{eqnarray}
\Vert \{\psi(t_1), F(A, B)\}\Vert
=
\int_{\omega_{t_1}\in \Omega}\mu_{\psi(t_1)}({\rm d}\omega_{t_1})
f_{F(A, B)}(\omega_{t_1})
\end{eqnarray}
and
\begin{eqnarray}
\Vert \{\psi(t_2), F(A, B)\}\Vert
=
\int_{\omega_{t_2}\in \Omega}\mu_{\psi(t_2)}({\rm d}\omega_{t_2})
f_{F(A, B)}(\omega_{t_2}).
\end{eqnarray}
Hence we have
\begin{eqnarray}
&&\Vert \{\psi(t_1), F(A, B)\}\Vert\cdot\Vert \{\psi(t_2), F(A, B)\}\Vert
\nonumber\\
&&=
\int_{\omega_{t_1}\in \Omega}\mu_{\psi(t_1)}({\rm d}\omega_{t_1})
f_{F(A, B)}(\omega_{t_1})
\int_{\omega_{t_2}\in \Omega}\mu_{\psi(t_2)}({\rm d}\omega_{t_2})
f_{F(A, B)}(\omega_{t_2}).\label{Einstein}
\end{eqnarray}
Proposition $\forall t\in{\bf R}, ~\forall \psi(t)\in
T, ~\forall A, B\in O : \left[
\Vert \{\psi(t), F(A, B)\}\Vert=0
\right]$ implies that 
the left-hand-side of (\ref{Einstein}) is 0.
We derive the following proposition:
\begin{eqnarray}
\Vert \{\psi(t_1), F(A, B)\}\Vert\cdot\Vert \{\psi(t_2), F(A, B)\}\Vert=0.
\label{norm}
\end{eqnarray}
On the other hand, Proposition M implies that 
the right-hand-side of (\ref{Einstein}) is as
\begin{eqnarray}
&&\int_{\omega_{t_1}\in \Omega}\mu_{\psi(t_1)}({\rm d}\omega_{t_1})
f_{F(A, B)}(\omega_{t_1})
\int_{\omega_{t_2}\in \Omega}\mu_{\psi(t_2)}({\rm d}\omega_{t_2})
f_{F(A, B)}(\omega_{t_2})\nonumber\\
&&
=
\int_{\omega_{t_1}\in \Omega}\mu_{\psi(t_1)}({\rm d}\omega_{t_1})
\int_{\omega_{t_2}\in \Omega}\mu_{\psi(t_2)}({\rm d}\omega_{t_2})
f_{F(A, B)}(\omega_{t_1})
f_{F(A, B)}(\omega_{t_2})\nonumber\\
&&
\le
\int_{\omega_{t_1}\in \Omega}\mu_{\psi(t_1)}({\rm d}\omega_{t_1})
\int_{\omega_{t_2}\in \Omega}\mu_{\psi(t_2)}({\rm d}\omega_{t_2})
|f_{F(A, B)}(\omega_{t_1})
f_{F(A, B)}(\omega_{t_2})|\nonumber\\
&&
=
\int_{\omega_{t_1}\in \Omega}\mu_{\psi(t_1)}({\rm d}\omega_{t_1})
\int_{\omega_{t_2}\in \Omega}\mu_{\psi(t_2)}({\rm d}\omega_{t_2})
=1.\label{Einstein1}
\end{eqnarray}
We use the following fact
\begin{eqnarray}
|f_{F(A, B)}(\omega_{t_1})\cdot
f_{F(A, B)}(\omega_{t_2})|=1.
\end{eqnarray}
Proposition L implies
\begin{eqnarray}
&&\{\omega_{t_1}|\omega_{t_1}\in\Omega\wedge
f_{F(A, B)}(\omega_{t_1})= 1\}=
\{\omega_{t_2}|\omega_{t_2}\in\Omega\wedge
f_{F(A, B)}(\omega_{t_2})= 1\},
\nonumber\\
&&\{\omega_{t_1}|\omega_{t_1}\in\Omega\wedge 
f_{F(A, B)}(\omega_{t_1})= -1\}
=
\{\omega_{t_2}|\omega_{t_2}\in\Omega\wedge 
f_{F(A, B)}(\omega_{t_2})= -1\}.\nonumber\\
\end{eqnarray}
The inequality (\ref{Einstein1}) is saturated since
\begin{eqnarray}
&&\{\omega_{t_1}|\omega_{t_1}\in\Omega\wedge
f_{F(A, B)}(\omega_{t_1})= 1\}=
\{\omega_{t_2}|\omega_{t_2}\in\Omega\wedge
f_{F(A, B)}(\omega_{t_2})= 1\},
\nonumber\\
&&\{\omega_{t_1}|\omega_{t_1}\in\Omega\wedge 
f_{F(A, B)}(\omega_{t_1})= -1\}
=
\{\omega_{t_2}|\omega_{t_2}\in\Omega\wedge 
f_{F(A, B)}(\omega_{t_2})= -1\}.\nonumber\\
\end{eqnarray}
Hence we derive the following proposition
if we assign the truth value ``1'' for
Proposition R, Proposition D, Proposition L, and Proposition M, simultaneously
\begin{eqnarray}
\Vert \{\psi(t_1), F(A, B)\}\Vert\cdot\Vert \{\psi(t_2), F(A, B)\}\Vert
=1.\label{M}
\end{eqnarray}

We do not assign the truth value ``1'' for two propositions
(\ref{norm}) and (\ref{M}), simultaneously.
We are in the contradiction.

We do not accept the following five propositions, simultaneously.
\begin{enumerate}
\item Proposition R
\item Proposition D
\item Proposition L
\item Proposition M
\item $\forall t\in{\bf R}, ~\forall \psi(t)\in
T, ~\forall A, B\in O : 
\left[
\Vert \{\psi(t), F(A, B)\}\Vert=0
\right]$.
\end{enumerate}
This is true for every time $t$, every state $\psi(t)$,
and every pair of observables $A$ and $B$.
Thus we have
\begin{eqnarray}
&&\forall t\in{\bf R}, ~\forall \psi(t)\in
 T, ~\forall A, B\in O : \nonumber\\
&&\left[
{\rm R}\wedge{\rm D}\wedge{\rm L}\wedge{\rm M}
\wedge 
\left[
\Vert \{\psi(t), F(A, B)\}\Vert=0
\right]
\right]
\Rightarrow 
\bot.
\end{eqnarray}
QED.

\section{Algebraic structure of 
observables and a local realistic theory}\label{MFinee}

In this section, we would like to infer main suggestion of this paper.

Let $B(\psi(t))$ be the symbol of the following proposition:

{\bf Proposition:} $B(\psi(t))$

Proposition R,
Proposition D, 
Proposition L, and
Proposition M
hold when the system is described by a state $\psi(t)$ in a time $t$.


\subsection{The Bell theorem and 
a commuting observables theory}

We might consider the following proposition:

{\bf Proposition:} C ({\it a local realistic theory 
implies commutativity of all observables}),

{\it A local realistic theory 
for all experimental events implies 
that an observable $F(A, B)$ 
is ${\bf 0}$, which represents the null observable,
for every pair of observables $A$ and $B$.
}

We would like to formulate Proposition C as the following proposition:
\begin{eqnarray}
\forall t\in{\bf R}, ~\forall \psi(t)\in T, ~\forall A, B\in O : 
\left[B(\psi(t))\Rightarrow  
\Vert\{\psi(t), F(A, B)\}\Vert=0\right].\label{Mfine}
\end{eqnarray}
The proposition (\ref{Mfine}) would be equivalent to the following proposition:
\begin{eqnarray}
\forall t\in{\bf R}, ~\forall \psi(t)\in T, ~\forall A, B\in O : 
\left[\overline{B(\psi(t))}\vee 
\Vert\{\psi(t), F(A, B)\}\Vert=0 \right].\label{Mfine2}
\end{eqnarray}
Here, the upper bar is for complements.

\begin{table}
\begin{tabular}{c | c | c }
$A$  &   $B$     &  $A\vee B$          \\ 
\hline
1 & 1       & 1                        \\
1 & 0       & 1                        \\
0 & 1       & 1                        \\
0 & 0       & 0                        
\end{tabular}
\caption{Truth Value Table : $A$ implies a proposition. 
$B$ implies a proposition. $A\vee B$ implies a proposition of 
disjunction of $A$ and $B$.}
\label{TRUTH}
\end{table}

\subsubsection{Case 1}

The Bell theorem is consistent with the negation of 
Proposition C as shown below.

If we assign the truth value ``1'' for the Bell theorem
and we assign the truth value ``0'' for Proposition C,
then we have to assign the truth value ``0'' for 
the following proposition:
\begin{eqnarray}
\forall t\in{\bf R}, ~\forall \psi(t)\in T : 
\overline{B(\psi(t))}.
\end{eqnarray}
From Truth Value Table \ref{TRUTH} and 
the fact that we assign the truth value ``0'' for Proposition C, 
we have to assign the truth value ``0'' for 
the following proposition: 
\begin{eqnarray}
\forall t\in{\bf R}, ~\forall \psi(t)\in T, ~\forall A, B\in O : 
\Vert\{\psi(t), F(A, B)\}\Vert=0.
\end{eqnarray}
From Truth Value Table \ref{TRUTH}, 
we have to assign the truth value ``0'' for 
the following proposition: 
\begin{eqnarray}
&&\forall t\in{\bf R}, ~\forall \psi(t)\in
 T, ~\forall A, B\in O : 
\nonumber\\
&&\left[
\overline{B(\psi(t))}
\vee
\left[
\Vert\{\psi(t), F(A, B)\}\Vert=0
\right]
\right].
\end{eqnarray}

Hence, we might say the following proposition:

{\bf Proposition 1:} 

The truth value for the following proposition is ``1'':
\begin{eqnarray}
&&[{\rm the\ Bell\ theorem}]\wedge
\overline{\rm Proposition\ C}.
\nonumber\\
\end{eqnarray}

Hence we would establish the desired consistency.


\subsubsection{Case 2}

The negation of the Bell theorem is consistent with 
Proposition C as shown below.

If we assign the truth value ``0'' for the Bell theorem
and we assign the truth value ``1'' for Proposition C,
then we have to assign the truth value ``1'' for 
the following proposition:
\begin{eqnarray}
\forall t\in{\bf R}, ~\forall \psi(t)\in T : 
\overline{B(\psi(t))}.
\end{eqnarray}
From Truth Value Table \ref{TRUTH} and 
the fact that we assign the truth value ``1'' for Proposition C, 
we may assign the truth value ``0'' or ``1'' for 
the following proposition: 
\begin{eqnarray}
\forall t\in{\bf R}, ~\forall \psi(t)\in
 T, ~\forall A, B\in O : 
\Vert\{\psi(t), F(A, B)\}\Vert=0.
\end{eqnarray}
From 
Truth Value Table \ref{TRUTH}, we have to assign the truth value ``1'' for 
the following proposition: 
\begin{eqnarray}
&&\forall t\in{\bf R}, ~\forall \psi(t)\in 
T, ~\forall A, B\in O : \nonumber\\
&&\left[
\overline{B(\psi(t))}
\vee
\left[
\Vert\{\psi(t), F(A, B)\}\Vert=0
\right]
\right].
\end{eqnarray}

Hence, we might say the following proposition:

{\bf Proposition 2:}

The truth value for the following proposition is ``1'':
\begin{eqnarray}
&&\overline{[{\rm the\ Bell\ theorem}]}\wedge
{\rm Proposition\ C}.
\nonumber\\
\end{eqnarray}

Hence we would establish the desired consistency.

\subsubsection{Case 3}

{\it The Bell theorem is consistent with  
Proposition C as shown below.}

If we assign the truth value ``1'' for the Bell theorem
and we assign the truth value ``1'' for Proposition C,
then we have to assign the truth value ``0'' for 
the following proposition:
\begin{eqnarray}
\forall t\in{\bf R}, ~\forall \psi(t)\in T : 
\overline{B(\psi(t))}.
\end{eqnarray}
From Truth Value Table \ref{TRUTH} and 
the fact that we assign the truth value ``1'' for Proposition C, 
we have to assign the truth value ``1'' for 
the following proposition: 
\begin{eqnarray}
\forall t\in{\bf R}, ~\forall \psi(t)\in T, ~\forall A, B\in O : 
\Vert\{\psi(t), F(A, B)\}\Vert=0.
\end{eqnarray}
From Truth Value Table \ref{TRUTH}, 
we have to assign the truth value ``1'' for 
the following proposition: 
\begin{eqnarray}
&&\forall t\in{\bf R}, ~\forall \psi(t)\in
 T, ~\forall A, B\in O : 
\nonumber\\
&&\left[
\overline{B(\psi(t))}
\vee
\left[
\Vert\{\psi(t), F(A, B)\}\Vert=0
\right]
\right].
\end{eqnarray}

Hence, we might say the following proposition:

{\bf Proposition 3:} 

The truth value for the following proposition is ``1'':
\begin{eqnarray}
&&[{\rm the\ Bell\ theorem}]\wedge
{\rm Proposition\ C}.
\nonumber\\
\end{eqnarray}

Hence we would establish the desired consistency.

\subsection{Conjecture}

Thinking of and estimating the validity of the Bell theorem
that would be either correct or not,
we might suggest the following statement:

{\bf Conjecture:}
{\it We conjecture that a realistic theory of the Bell type 
(a local realistic theory)
for events would not imply 
the commutative algebraic structure into 
the set of all observables
if all experimental events
would be reproduced by a local realistic theory.
}

\section{Bell locality
and a realistic theory of the Kochen-Specker type}\label{exam}

In this section, it would be suggested 
that a violation of Bell locality (Bell locality is defined by Proposition L) 
would be derived 
within a realistic theory of the Kochen-Specker (KS) type for quantum events.
A violation of Bell locality is as
\begin{eqnarray}
\omega_{t_1}\neq \omega_{t_2}
\end{eqnarray}
for ${t_1}\neq{t_2}$.

It is suggested \cite{bib:Nagata2} that 
a realistic theory of the Bell type ($\omega_{t_1}= \omega_{t_2}$)
would violate a realistic theory of the KS type 
(noncommutativity would appear)
in a finite-dimensional Hilbert space formalism of the quantum theory.
An inequality valid for 
the KS condition for a finite number of observables is used.
The quantum predictions produced 
by an uncorrelated pure state would violate the inequality 
(within a finite-dimensional space).

It is suggested \cite{bib:Nagata1} that 
a violation of some entanglement 
witness inequality would imply noncommutativity 
(within a finite-dimensional space).

We might consider that a violation of 
some inequality would imply noncommutativity.
This conjecture might be true.
If so, we might ask by what origin such an inequality is derived is.

By the way, we would like to review the functional rule (the KS condition).
Let $g$ be any function. Let $v$ be $v\in O$.
The functional rule is 
\begin{eqnarray}
g(f_{v}(\omega_t))=f_{g(v)}(\omega_t)\label{KS}
\end{eqnarray}
for every $t\in {\bf R}$.
If we would accept the condition (\ref{KS}), 
all quantum observables in 
a finite-dimensional Hilbert space $H$ 
would commute simultaneously.

The inequality in question 
would be originated from 
\begin{enumerate}
\item the functional rule (the KS condition)
\item the quantum relation 
$\sigma_z^1\sigma_z^2=-\sigma_y^1\sigma_y^2\sigma_x^1\sigma_x^2$
\end{enumerate}
since the discussion of the inequality in Ref.~\cite{bib:Nagata2} 
uses these conditions.
This might be reasonable in the following sense:
`A violation of an inequality derived by commutative condition 
implies noncommutativity'.

We notice Peres no-go theorem \cite{bib:Peres2}.
We see that the inequality in question relies on Peres no-go theorem.
There is an elegant key in 
Peres no-go theorem, which suggests a violation of Bell locality within 
a realistic theory of the KS type.

We would like to 
fix the value as $f_{\sigma_z^1\sigma_z^2}(\omega_t)=\zeta(\pm 1)$ in 
two spin-$\frac{1}{2}$ particles in any quantum state 
for every $t\in{\bf R}$.
Then the values of $f_{\sigma_x^1\sigma_x^2}(\omega_t)$ and 
$f_{\sigma_y^1\sigma_y^2}(\omega_t)$ are 
necessarily entangled each other in
the sense that 
the value of one of them is immediately 
determined when we assign a value to other.

It would be thinkable that such an 
entanglement would have nothing to do with the validity of 
a violation of Bell locality ($\omega_{t_1}\neq \omega_{t_2}$).
We would like to assume the validity of 
a violation of Bell locality ($\omega_{t_1}\neq \omega_{t_2}$).
Then the values of 
$f_{\sigma_x^1\sigma_x^2}(\omega_{t_1})$,
$f_{\sigma_y^1\sigma_y^2}(\omega_{t_1})$, and
$f_{\sigma_z^1\sigma_z^2}(\omega_{t_1})$
are independent of each other.
The values of 
$f_{\sigma_x^1\sigma_x^2}(\omega_{t_2})$,
$f_{\sigma_y^1\sigma_y^2}(\omega_{t_2})$, and
$f_{\sigma_z^1\sigma_z^2}(\omega_{t_2})$
are also independent of each other.

We might call such an entanglement as {\it the KS nonlocality}.
The KS nonlocality of value assignment appears when we would like to accept 
the KS condition in 
a finite-dimensional Hilbert space.
The KS nonlocality depends on 
algebraic structure of observables.
The KS nonlocality is 
independent of the state of the system under study.
The KS nonlocality is 
independent of time $t$.

In contrast, Bell nonlocality ($\omega_{t_1}\neq \omega_{t_2}$)
itself would not certify 
such a nonlocality of value assignment.
Bell nonlocality does not depend on 
algebraic structure of observables.
Bell nonlocality depends on 
the state of the system.
Bell nonlocality depends on 
time $t$.

We would like to hope that our discussion might 
infer an example which suggests
a violation of Bell locality 
(a violation of classical algebraic structure 
(physically speaking, a violation of time symmetry)) 
within 
a realistic theory of the KS type 
(classical algebraic structure
(physically speaking, Newton's theory)).

\section{Conclusions}\label{conclusion}

In conclusion, 
we have studied
the relation between a local realistic theory 
and commutativity of observables in 
a finite-dimensional space.
We have conjectured that a realistic theory of the Bell type 
(a local realistic theory)
for events would not imply 
the commutative algebraic structure into 
the set of all observables
if all experimental events
would be reproduced by a local realistic theory.
We have suggested that a violation of Bell locality would be derived 
within a realistic theory of the Kochen-Specker type 
within quantum events.

We would like to hope that 
our discussion might provide further information for 
people who further consider the open problem.

\section*{Acknowledgments}

The author would like to 
thank Professor Marek \.Zukowski for valuable discussions.
The author would like to 
acknowledge Professor Tadao Nakamura for his helpful much 
education.
This work has been supported by Frontier Basic Research Programs at
KAIST and K.N. is supported by a BK21 research grant.

\end{document}